\documentstyle[epsfig,fleqn,twoside,gc-98]{article}

\def\la{\lower.5ex\hbox{$\; \buildrel < \over \sim \;$}}
\def\ga{\lower.5ex\hbox{$\; \buildrel > \over \sim \;$}}
\heads{Safonova M.V. and Lohiya D.}
      {Gravitational lensing on gravity balls}

\begin{document}
 
\mathindent 0in
\twocolumn[
\Arthead{00}{1}{10}

\Title{Gravity balls in induced gravity model---
                 'gravitational lens' effects}
\Author{M. V. Safonova \foom 1             
              and D. Lohiya \foom 2}    
          {Dept. Physics \& Astrophysics, \\University of Delhi,
Delhi-7, India }

\Abstract
{In non-minimally coupled effective gravity theories one can have
non-topological solitonic solutions. A typical solution is a spherical
region with $G_{\it eff}=0$ outside and having the canonical Newtonian
value inside. Such a spherical domain (gravity-ball) is characterized
by an effective index of refraction which causes bending of light
incident on it. The gravity ball thus acts as a spherical lens. We
consider the gravity ball to be of a size of a typical cluster of
galaxies and show that even empty (without matter) gravity ball can
produce arc-like images of the background source galaxy. In the case of
background random galaxy field the ball produces distortions (`shear')
of that field. We also obtained constraints on the size of the large
gravity ball, which can be inferred from the existing observations of
clusters with arcs.}

]  

\email 1 {rita@iucaa.ernet.in}
\email 2 {dlohiya@ducos.ernet.in}

\section{Introduction}
One of the remarkable results of field theory is the existence of
stable classical states of a field with non-vanishing energy. Such
fields include topological solitons, domain walls, strings and
monopoles. In addition to these field configurations, which are
stabilized by their topological properties, non-topological solitons
(NTS) produced by scalar fields have appeared in the literature and
their relevance to cosmology has been assessed (for example, \cite{1}).
Variations on this theme include cosmic neutrino balls \cite{2},
\cite{3}, Q-balls \cite{4} and soliton stars \cite{5}. NTS are
rendered stable by the existence of a conserved Noether charge carried
by the fields confined to a finite region of space. The theory
essentially contains an additive quantum number $N$ carried either by a
spin-1/2 field $\psi$ (for example, fermion number), or a spin-0
complex field $\varphi$. In addition there is a scalar field $\sigma$
residing at $\sigma=0$ in the normal (true) vacuum state and at
$\sigma=\sigma_0$ in the local (false) vacuum state. The soliton
contains the interior in which $\sigma \approx \sigma_0$, a shell of
width $\approx \mu^{-1}$ over which $\sigma$ changes from $\sigma_0$ to
$0$ ($\mu$ being the mass associated with the scalar field), and
finally the exterior that is the true vacuum. The $N$-carrying field
$\psi$, or $\varphi$, is confined to the interior where it is
effectively massless at the local minimum of the potential,
$\sigma=\sigma_0$ (at the global minumum, $\sigma=0$ the field ($\psi$
or $\varphi$) has a non-vanishing mass, $m$). This leads to a stable
configuration of massless particles trapped inside a region with
$\sigma=\sigma_0$ separated from the true vacuum $\sigma=0$ by a wall
of thickness $\approx \mu^{-1}$. Confinement occurs for all particle
states that are not on-shell in the exterior region. The NTS will be
stable as long as the kinetic energy $E_k$ of the particles inside the
bag is less than $E_{free}$---the minimum on-shell energy in the
exterior region. In the scenario of formation of the NTSs upon the
spontaneous breaking of discrete symmetry, two regions of different
vacua (true and false) are separated by a domain wall. A bubble of true
vacuum forms in a domain trapped in a false ground state. The false
state collapses due to the surface tension of the boundary. The
collapse is inhibited, however, by the pressure exerted by (for
example) massless fermions, which cannot leave the interior of the
bubble since their energy is less than their mass outside the bubble.
This was explicitely demonstrated by Lee and Wick \cite{6},\cite{7}.
 
The role of scalar fields in effective gravity models stems from the
classical work of Brans and Dicke. In these approaches the
gravitational action is induced by a coupling of the scalar curvature
with a function of a scalar field. Lee and Wick's results can be
carried over to a curved spacetime with a scalar curvature
non-minimally coupled to $\sigma$ in the class of theories described by
the effective action

\begin{eqnarray} 
\lefteqn{S = \int \sqrt{-g} \,d^4x \left [ U( \sigma)R 
+ \frac{1}{2} (\partial_{ \mu} \sigma)^2 -V( \sigma) +\right.} \nonumber \\
& & \left. \tilde{U}(\sigma)\left(m \bar{\psi} \psi + \bar{ \psi} \gamma^{
  \mu}\partial_{ \mu} \psi \right)\right ] 
\end{eqnarray} 
Establishing $\sigma_{in}$ and $\sigma_{out}$ as the interior and exterior
values of $\sigma$, the
effective gravitational constant would be given by
$$
G_{\it eff}^{\it in} = [U(\sigma_{in}]^{-1}
$$
and
$$
G_{\it eff}^{\it out} = [U(\sigma_{out}]^{-1}
$$
If $U$ goes to infinity at some point, defined as $\sigma = 0$ without
loss of generality, the theory gives rise to a solution with a spatial
variation of the effective gravitational constant, $G_{\it eff}^{\it
out}=0$ and $G_{\it eff}^{\it in}=const$. This would be a generic
feature of a Lee-Wick solution in which the scalar field is
non-minimally coupled to the scalar curvature. It would give rise to a
stable ``ball of gravity''. In the present paper, we examine this
special kind of NTS solutions referred to as G-balls \cite{8}.

A gravity ball is characterised by an effective refractive index in its
interior. This would cause bending of light incident on it. The G-ball
thus acts as a spherical lens. Gravitational lensing (GL) is a powerful
tool in exploring the Universe. The topic has presently matured to the
stage where GL is applied by astronomers for a variety of purposes and,
in particular, it is providing an exciting new probe of detecting
exotic objects in the Universe (may be described by the matter fields
which are not detected yet) as well as of testing alternative theories
of gravity. Proposals have been made to discover cosmic strings
\cite{9}, boson stars \cite{10} and neutralino stars \cite{11},\cite{12} 
through their gravitational lensing effects. There is no compelling
evidence that any of the observed GL systems are due to these objects,
however, it is essential to develop new lens models with objects which
are though not yet observed, but not forbidden on the theoretical
grounds. In the present paper we investigate gravitational lensing of
an empty gravity ball situated at a cosmological distance. The
lens---gravity ball---has interesting features which are not shared by
other known lenses. The lensing effect is radically different---rays
with impact parameter greater than the radius of the ball are not
deflected.

The paper is organised as follows. In the Section 2 we give the general
formalism of the problem for a gravity ball, in the Section 3 we
discuss the effect of GL by gravity ball, namely, the number of images,
magnification, etc. Finally in the last section we give the
conclusions.
 
\section{General formulations}
\subsection*{Gravity ball as a NTS solution}
Here we will discuss the general formalism of the problem for a
spherically symmetric system consisting of fermion field $\psi$, scalar
field $\sigma$ and gravitational field $g_{\mu \nu}$. We follow the
theory presented by Sethi and Lohiya \cite{8}. Gravity balls are the
NTS solutions of the field equations arising from (1)
 
\begin{eqnarray*}
\lefteqn{U(\sigma) [R^{\mu \nu} - \frac{1}{2} g^{\mu \nu} R] = -\frac{1}{2} 
\left [T^{\mu \nu}_{\omega} +
T^{\mu \nu}_{\sigma} + \right.}\\
 & & \left. T^{\mu \nu}_{\sigma,\psi} + T^{\mu \nu}_{\psi} + 
2U(\sigma)^{;\mu;\nu} - 2 g^{\mu \nu} U(\sigma)^{;\lambda}_{;\lambda} 
\right ]  
\end{eqnarray*}
\begin{equation}
g^{\mu \nu} \sigma_{;\mu;\nu} + \frac{\partial V}{\partial \sigma} - R
\frac{\partial U}{\partial \sigma} = 0
\end{equation}
Here $T^{\mu \nu}_{\sigma}$, $T^{\mu \nu}_{\psi}$,
$T^{\mu\nu}_{\sigma,\psi}$ and $T^{\mu \nu}_{\omega}$ are energy
momentum tensors constructed from action for the scalar field, the
fermion field, together with its Higgs coupling to $\sigma$, and the
rest of the matter fields, respectively. We consider an NTS with the
scalar field held to a value $\sigma_0$ in the interior and making a
fast transition to $\sigma=0$ outside a thin shell. Thus, we have
essentially three regions, interior of a soliton ($r<R_0$), a shell of
thickness $\sim \mu^{-1}$ and surface energy density $s \approx \mu
\sigma_0^2/6$, and exterior ($r>R_0$). The total energy of a NTS has
contributions from: (1) the surface tension energy $E_s \approx s
R_0^2\,$; (2) the energy of the fermions $E_f \approx N^{4/3}/R_0$ and;
(3) the volume energy $E_V \approx V(\sigma_{\it in})R_0^3$. For the
degenerate case $V(\sigma_{\it in})=0$, a NTS has total mass
constrained by the stability against gravitational collapse to a value
determined by the surface tension $s$. The soliton mass, obtained by
minimizing the total energy, is $M=12 \pi s R_0^2$. For $s \sim
(Mev)^3$ and $N \approx 10^{75}$, the size of the NTS is of the order
of tens of kiloparces, while it is still away from the Schwarzschild
bound. For configurations with $R$ much greater than the Schwarzschild
radius, the effects of gravity can be treated as a small perturbation.
Thus, the form of the metric for the NTS satisfies a "weak field
approximation".

\paragraph{A. Interior: $r < R_0 + {\cal O}(\mu^{-1})$.}
The metric inside is described as
\begin{equation}
ds^2 = e^{2u(r)}\,dt^2 - e^ {2v(r)}\,dr^2 - r^2 [d\theta^2 +
\sin^2\theta\,d\varphi^2]
\end{equation}
The interior metric has one specific solution:
$$
v = -\frac{\hat{C} r^2}{6}
$$
and
\begin{equation} 
u = u_0 + \frac{r^2}{2} \left[ \frac{\tilde{C}}{2} + \frac{\hat{C}}{3} \right]
\end{equation}
with $\tilde{C}$ and $\hat{C}$---constants depending on the fermionic
energy inside the soliton. In the weak field approximation used here we
consider the interior with
\begin{equation}
ds^2 \approx e^{2u_0}\,dt^2 - dr^2 - r^2[d\theta^2 + \sin^2{\theta\,
d\varphi^2}]
\end{equation}

\paragraph{B. Exterior: $r > R_0 + {\cal O}(\mu^{-1})$.}
In the exterior region we have essentially Minkowskian metric
\begin{equation}
ds^2 = dt^2 - dr^2 - r^2 [d\theta^2 +\sin^2{\theta\,d\varphi}^2]
\end{equation}

To be consistent with the observations we will see that the $u_0$ has
to be small negative constant. The propagation of light inside the
G-ball is equivalently described by using the Fermat principle with the
effective refraction index $n_{\it eff}$ inside the ball given by
\begin{equation}
n_{\it eff} = 1 - u_0
\end{equation}
This gives straight trajectories of light rays inside the ball. But
since outside the ball $n_{\it eff}=1$, we see that G-ball behaves as a
spherical lens (deflection of light occurs only at the boundaries).

\section{Lens Model for a G-ball and Lensing Properties}
Throughout the paper, we employ the conventions of many articles, reviews and
books on gravitational lensing (see, for example, \cite{13}, \cite{14}) and which
are essentially illustrated in the Fig.1.

\begin{figure}[ht]
\centerline{
\epsfig{figure=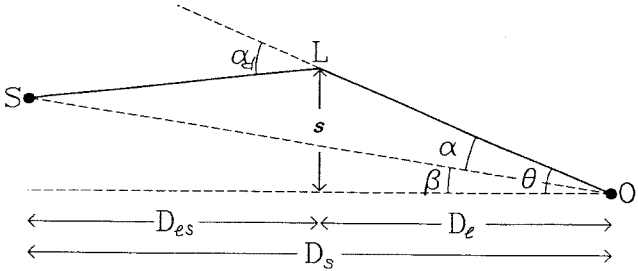,width=0.5\textwidth}}
\caption{Basic geometry of gravitational lensing. The light ray propagates from
the source $S$ to the observer $O$, passing the lens $L$ at transverse
distance $s$. Angle $\alpha_d$ is the angle of deflection. The angular
separations of the source and the image from the optic axis as seen by
the observer are $\beta$ and $\theta$, respectively. The distances
between the observer and the source, the observer and the lens, the
lens and the source are $D_s$, $D_l$ and $D_{ls}$, respectively.}
\end{figure} 
 
The positions of the source and the image are related through the {\it
lens equation} 
\begin{equation}
 \vec{\beta}=\vec{\theta} -
\frac{D_{ls}}{D_s}\vec{\alpha_d} 
\end{equation} 
where we employ angular diameter distances. We restrict our discussion
to a cosmological model which is a variant of a Milne universe \cite{8}
with a scale factor $R(t)=t$ and cosmological parameters $\Omega_0=0$,
$\Omega_{\Lambda}=0$. In this cosmology angular diameter distances are
given by

\begin{equation}
{\cal D}_l = \frac{c z_l}{2 H_0}\frac{(2+z_l)}{(1+z_l)^2}
\end{equation}
\begin{equation}
{\cal D}_s = \frac{c z_s}{2 H_0}\frac{(2+z_s)}{(1+z_s)^2}  
\end{equation}
\begin{equation}
{\cal D}_{ls} = \frac{c z_{ls}}{2H_0(1+z_l)}\frac{(2+z_{ls})}{(1+z_{ls})^2} 
\end{equation}
with 
\begin{equation}
z_{ls} = \frac{z_s-z_l}{1+z_l}
\end{equation}
and $H_0=100\,h$ km s$^{-1}$ Mpc$^{-1}$ is the Hubble constant. 
Subscripts $l$, $s$ and $ls$ stand for lens, source and lens-source, 
correspondingly.

In our subsequent development we also utilize the following
assumptions, mostly related to the modelling of the deflector. We
consider the gravity ball to be of the size of a typical cluster of
galaxies at the cosmological distance and embedded in the empty
space---region with no matter concentrations close to it. The deflector
is transparent. We also assume a thin lens approximation since all
deflection occurs within $\Delta z \le \pm R_0$; the extent of the
deflector is thus taken to be small compared with its distance from
both the observer and the source (say, distances from us to the cluster
at $z=0.3$ and to the source at $z=1$ are $\sim 1\,\,Gpc$ and $\sim
2\,\, Gpc$, respectively, while $R_0 \approx 0.5$ to $1$ Mpc).  We
regard deflection angles to be small. We also assume that the positions
of the source, lens, and observer are stationary with respect to
comoving coordinates.

The magnification of images is given by
$$
\mu= \left|det\frac{\partial\vec{\beta}}{\partial\vec{\theta}}\right|^{-1} 
$$
which in axisymmetric case goes to
\begin{equation}
\mu = \left( \frac{\beta \,d\beta}{\theta\,d \theta} \right)^{-1}
\end{equation}
The tangential and radial critical curves folow from the singularities of the
tangential and radial magnification
\begin{equation}
\mu_t \equiv \left (\frac{\beta}{\theta} \right)^{-1}
\end{equation}

\begin{equation}
\mu_r \equiv \left( \frac{d\beta}{d\theta} \right)^{-1}
\end{equation}

\paragraph {Lens equation.}

\begin{figure}[ht]
\centerline{
\epsfig{figure=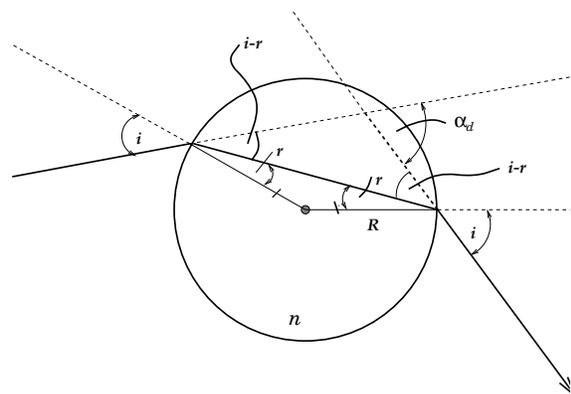,width=0.5\textwidth}}
\caption{$R_{ball}$---radius of the 
ball, $i$---angle of incidence, $r$---angle of refraction, $\alpha_d$---
deflection angle, $n$---ratio of the refractive index inside the ball to the 
refractive index outside; $n>1$.}
\end{figure}

In the Fig. 2 we present the geometry of G-ball as a lens. 
From the Fig.2 and Snell's law the deflection angle $\alpha_d$ is
\begin{equation}
\alpha_d = 2 \left[i - \sin^{-1} \left( \frac{1}{n}\sin i \right) \right]
\end{equation}
Since we consider all angles to be small, the following approximations 
are valid. Defining the quantity
\begin{equation}
\theta_C = \frac{R_{\it ball}}{D_l}
\end{equation}
which is the radius of the region inside which the refraction occurs,
we obtain the the expression for the deflection angle $\alpha_d$ 
\begin{equation}
\alpha_d = 2 \left[ \sin^{-1} \left( \frac{\theta}{\theta_C} \right) -
\sin^{-1} \left( \frac{\theta}{n \theta_C} \right) \right]
\end{equation}
Plugging (18) into the general lens equation (8) we obtain lens
equation for the gravity ball  
$$
\vec{\beta} = \vec{\theta} - \frac{2 D_{ls}}{D_s} \frac{\vec{\theta}}
{\theta} \left [ \sin^{-1} \left( \frac{\theta}{\theta_C} \right) -
\sin^{-1} \left( \frac{\theta}{n \theta_C} \right) \right ]  
$$
\begin{equation}
\mbox{for $ 0 \le \theta \le \theta_C$}
\end{equation} 
and 
$$
\vec{\beta} = \vec{\theta} 
$$
\begin{equation}
\mbox{for $ \theta > \theta_C$}, 
\end{equation}
where $\theta \equiv |\vec{\theta}| = \sqrt{\theta_1^2+\theta_2^2}$ is the 
radial position of the image in the lens plane.

\paragraph{Multiple image diagram and conditions for multiple imaging.}
To illustrate the lensing properties described by equations (19,20) we
show in Figure 3 the lensing curve for the gravity ball with parameters
$n=1.0005$, $\theta_C=5'$ and for $z_{\it source}=1$, $z_{\it
lens}=0.3$. The intersections of the lines $\beta = const$ with the
curve given by equation (19) give the solutions to the lensing
equation. The source at $\beta_2$ lies on a caustic point and
$\theta_r$ is the radius of the radial critical curve; the source at
$\beta_1$ has three images; and point $\beta=0$ produces a
ring---`Einstein ring'--- with the critical radius $\theta=\theta_t$
and an image at $\theta=0$. The function $\beta(\theta)$ is continuous
except at $\theta_C$ due to the edge of the ball.

\begin{figure}[ht]
\centerline{
\epsfig{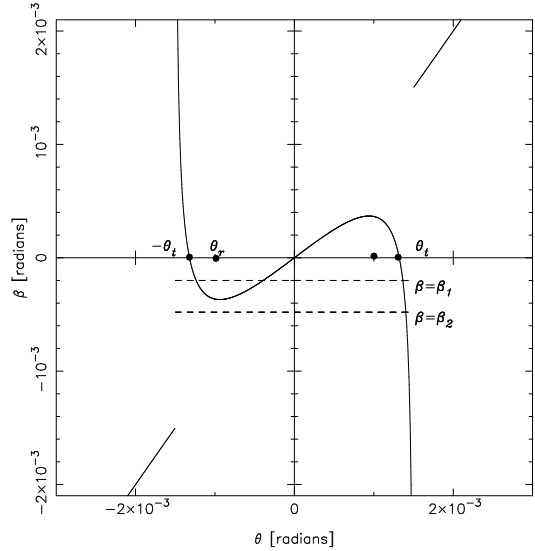}}
\caption{Solution to the lensing equation by a gravity ball. The solid curve
represents the lensing equation curve with $n=1.0006$, $\theta_C=5'$,
$z_{source}=1$, $z_{lens}=0.3$, together with lines $\beta =\beta_i$
(dashed lines) for various source positions $\beta_i$. The
intersections of the lines $\beta = const$ with the lensing equation
curve give the number and positions of the lensed images. The source at
$\beta_2$ lies on a caustic point and $\theta_r$ is the radius of the
radial critical curve; source at $\beta_1$ has three images and point
$\beta=0$ produces a ring $\theta_C=\theta_t$; in addition, it has an
image at $\theta=0$. The function $\beta(\theta)$ is continuos except
at $\theta_C$ due to the edge of the ball.}
\end{figure}

Putting $\beta=0$ in (19) gives
\begin{equation}
\theta -   \frac{2D_{ls}}{D_s} \left( \sin^{-1}{\frac{\theta}{\theta_C}} -
\sin^{-1}{ \frac{\theta}{n \theta_C}} \right) = 0 
\end{equation}
Denoting $a = 2D_{ls}/D_s$
\begin{equation}
\theta = a \left( \sin^{-1}\frac{\theta}{\theta_C} -\sin^{-1}{ 
\frac{\theta}{n \theta_C}} \right)
\end{equation}
When the alignment of the source, lens and the observer is not perfect,
we see the features called arcs, which are the result of very strong
distortion of a background sources. Arcs roughly trace the Einstein
ring, so $\theta_{\it arc} \approx \theta_E$. Since radii of most
known arcs do not exceed $30''$ small angle approximation is valid and
we obtain

\begin{eqnarray}
\lefteqn{\theta \left\{1 -  \frac{a}{\theta_C} \sqrt{1-\left(\frac{\theta}{n \theta_C} 
\right)^2} + \right. }  \nonumber \\
 & & \left. \frac{a}{n \theta_C} \sqrt{1-\left(\frac{\theta}{\theta_C}\right)^2} 
\right \} = 0
\end{eqnarray}
One solution to this equation is trivial:
\begin{equation}
\theta = 0
\end{equation}
which corresponds to the image located at the
centre of the lens. To find other solutions we write the expression 
in the curly brackets in (23) as 
\begin{equation}
\theta = \theta_C \sqrt{1 - \frac{a^2 n^2}{4 \theta_C^2} \left 
(1 - \frac{\theta_C^2}{a^2} - \frac{1}{n^2} \right)^2}
\end{equation}
and in order to find the conditions for the Einstein ring
we rewrite it in the form
\begin{equation}
\theta_E = \theta_C \sqrt{1 - \frac{a^2 n^2}{4 \theta_C^2} \left 
(1 - \frac{\theta_C^2}{a^2} - \frac{1}{n^2} \right)^2}
\end{equation}
Assuming for simplicity $a = 1$ (the ball (lens) is half-way between the 
observer and the source) and using $n$ from other estimates \cite{8} 
to be from $1.001$ to $1.0001$ 
we obtain\newline 
for $n = 1.0001$
\begin{equation}
\theta_E \cong 0.98\,\theta_C
\end{equation}
for $n = 1.001$
\begin{equation}
\theta_E \cong 0.73\,\theta_C
\end{equation}
Here $\theta_E$ is non-trivial solution of the equation (23). This
leads to the appearence of the image in the form of a ring
with the radial size of $\sim 0.9$ to $0.7$ of the total size of the ball,
depending on $n$. To find the condition for the appearence of
multiple images we analyze (25). To obtain a physical solution we take
\begin{equation} 
1 - \frac{a^2 n^2}{4 \theta_C^2} \left (1 - \frac{\theta_C^2}{a} - 
\frac{1}{n^2} \right)^2 > 0
\end{equation}
The conditions for multiple imaging are
\begin{equation}
R_{\it ball} > 2 D_{\it eff} \left(1- \frac{1}{n} \right) 
\end{equation}
or expressed through $n$
\begin{equation}
1 < n < \frac{2 D_{\it eff}}{2 D_{\it eff} - R_{\it ball}} 
\end{equation}
where $D_{\it eff} = \frac{D_{ls} D_l}{D_s}$ is effective distance.

\paragraph{Magnification and critical curves.}
From lens equation for a G-ball (19) and equation (13) 
we obtain expression for the total magnification of the images   
\begin{eqnarray}
\lefteqn{\mu^{-1} = \left \{1 - \frac{a}{\theta} \left [ \sin^{-1}
\left( \frac{\theta}{\theta_C} \right) - \sin^{-1}\left (\frac{\theta}{n\,\theta_C} 
\right ) \right ] \right \} \times} \nonumber \\ 
 & & \left \{ 1 - a \left ( \frac{1}
{\sqrt{\theta_C^2-\theta^2}} - \frac{1}{\sqrt{n^2 \theta_C^2-
\theta^2}} \right ) \right \}
\end{eqnarray}
where first term represents $\mu_t$ and second term---$\mu_r$. In
Figure 4 we have plotted tangential magnification $\mu_t$ and radial
magnification $\mu_r$ against $\theta$ for a G-ball with refraction
index $n=1.001$. Singularities in these give the angular positions of
tangential critical curves and radial critical curves, respectively. In
the same Figure we also plotted total magnification $\mu$ vs.
$\theta$.
    
\begin{figure}[ht]
\centerline{
\epsfig{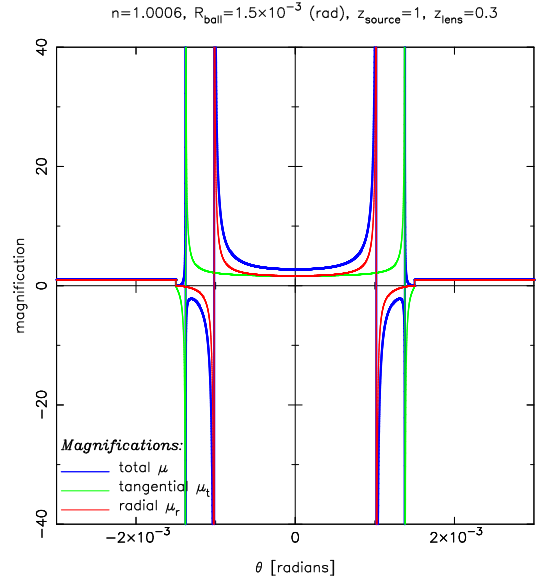}}
\caption{The magnifications: tangential $\mu_t$ denoted by green, radial $\mu_r$
denoted by red, and total $\mu$ is shown by blue line. Curves are
plotted as a function of the image positions $\theta$ for the
parameters of the lens: $n = 1.0006$, $\theta_C=5'$, $z_l = 0.3$, $z_s
=1$. The singularities of $\mu_r$ and $\mu_t$ give positions of the
tangential and radial critical curves, respectively.} 
\end{figure}

\paragraph{Limits on $n$ from observations.}
We can rewrite equation (26) in terms of variables $R_{\it ball}$ and 
$R_E$ to get an expression for $R_{\it ball}$ in terms of $n$ and a 
given $R_E$.
\begin{equation}
R_{\it ball}^2 = 4 D_{\it eff} \left [ 1+ \frac{1}{n^2} -
\frac{\sqrt{4 D_{\it eff}^2 -R_E^2}}{n D_{\it eff}} \right ]  
\end{equation}
Since $R_{\it ball} \ge R_E $, we get a lower bound on the value of
$R_{\it ball}$ if we know the radius of the Einstein's ring. It is
clear from the above equation that if the size of the G-ball is big
enough then a large enough $n$ can give us any desired radius for the
Einstein's ring. If we assume that all G-balls are of the same size
then from observations we can infer a lower bound on the radius and the
refractive index inside a ball by the following argument. The radius of
any gravity ball has to be larger then the radius of the largest
observable Einstein's ring. Given this size of the ball the refractive
index should be large enough to give a real Einstein's ring for every
other case. This situation is illustrated in the Figure 5, where
observational data for few clusters with giant arcs (details are
presented in the Table 1) are used with the assumption that the raduis
of the arc $\theta_{\it arc} \approx \theta_E$. Together with the curve
(33) for the cluster A370 we plotted the value of the radius $R_{\it
arc}$ of the A5 arc in that cluster (horizontal line), which is the
largest amongst presented clusters. It is clear that the refractive
index should be greater than the value where this line intersects the
$R_{\it ball}$--$n$ curve for this cluster. Assuming $R_{\it ball}$ to
be the same, we can see that in order to have an arc, a gravity ball
must have $n$ more than $n \approx 0.00056$. Thus, we obtained a lower
limit on $n$. Of course, our present calculations are for the case of
empty gravity ball. Matter concentrations in the centre of the ball
increase the deflection angle. We intend to explore in details this
analysis in our next~paper.

\begin{figure}[ht]
\centerline{
\epsfig{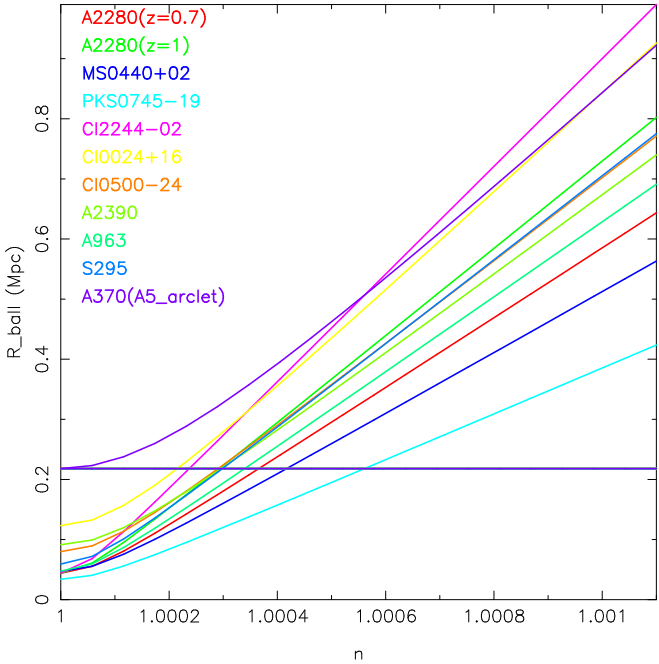}}
\caption{$R_{\it ball} $ vs. $n$ for different clusters. Clusters are
marked by different colours. Corresponding curves have the same colour as
the name of the cluster. Details are presented in Table 1. Horizontal 
line is the value of radius of the arc A5 in the cluster A370.} 
\end{figure}

\paragraph{Simulations.} 
To demonstrate how a G-ball at intermediate redshift gravitationally
distorts background sources we performed computer simulations and
presented the results in Figures 6 and 7. We simulated empty gravity
ball of a size of a typical galaxy cluster at the reshift of
$z_{l}=0.3$. We assumed for simplicity that all source galaxies are at
the same redshift $z_s=1$. The luminous area of each background galaxy
is taken to be a circular disk of radius $R$ with constant brightness.
In all our numerical computations we used Hubble parameter $h=1$. For
the simulations of G-ball lensing we use the algorithm described in
\cite{13}. The image configurations by a simulated gravity ball are
illustrated in Figure 6 by changing the relative positions between the
extended circular source and the lens. When the alignment of the source
on the optical axis is perfect, the ring image---Einstein
ring---appears (Fig.6a); with a small displacement we see two opposite
arcs approximately located at the Einstein radius (Fig.6e). When the
displacement becomes larger different features may appear: single
tangential arc Fig.6(b), radial arc (Fig. 6(d), two opposite images
(Fig. 6(c) and straight arc (Fig.6(f)). Figure 7 displays the disortion
the gravity ball produces on the field of randomly distributed
background sources. Figure 7(a) shows an unperturbed galaxy backgound
projected randomly in a field of $\sim 10' \times 10'$ at a redshift of
$z=1$. Fig. 7(b) illustrates the same field with an empty G-ball at the
centre. To compare with the lensing effect a cluster modelled as a
singular isothermal sphere produces on the background galaxies, we
presented in the Fig.7(c) the simulation taken from the \cite{15}. We
can see from the figures that an empty gravity ball can indeed act as a
strong lens and produce the images of arcs and arclets.

\begin{figure}[ht]
\centerline{
\epsfig{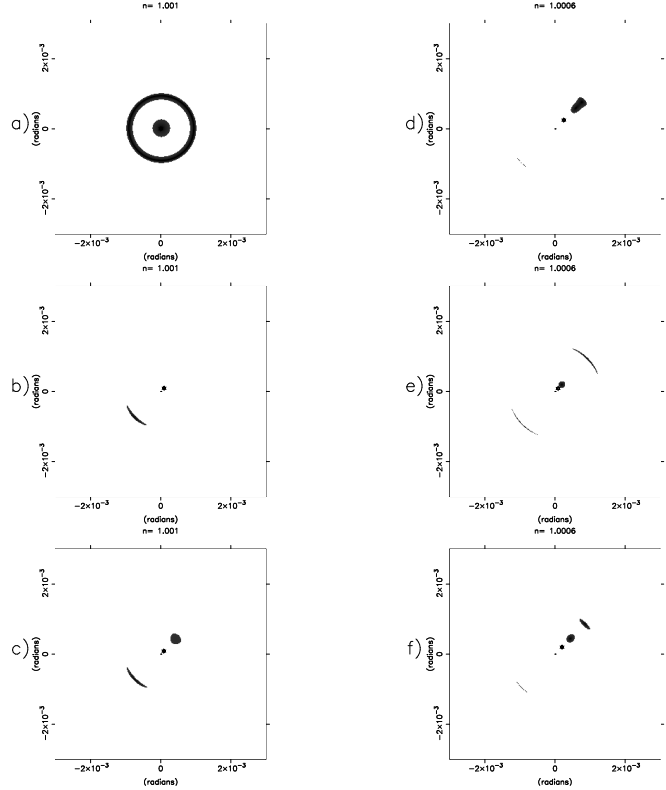}}
\caption{Illustration of six different imaging situations for an empty gravity
ball. $z_l=0.3$, $z_s =1$, $\theta_C=5'$. Depending on the source
position and $n$ the lens produces images of (a) ring, (b) single arc,
(c) arc and opposite image, (d) radial arc, (e) three images (with two
opposite arcs), (f) two images on one side (with a straight arc). Small
dot in the centers of the panels marks the center of the G-ball. Small
star marks the position of the source.}
\end{figure} 

\begin{figure}[ht]
\centerline{
\epsfig{figure=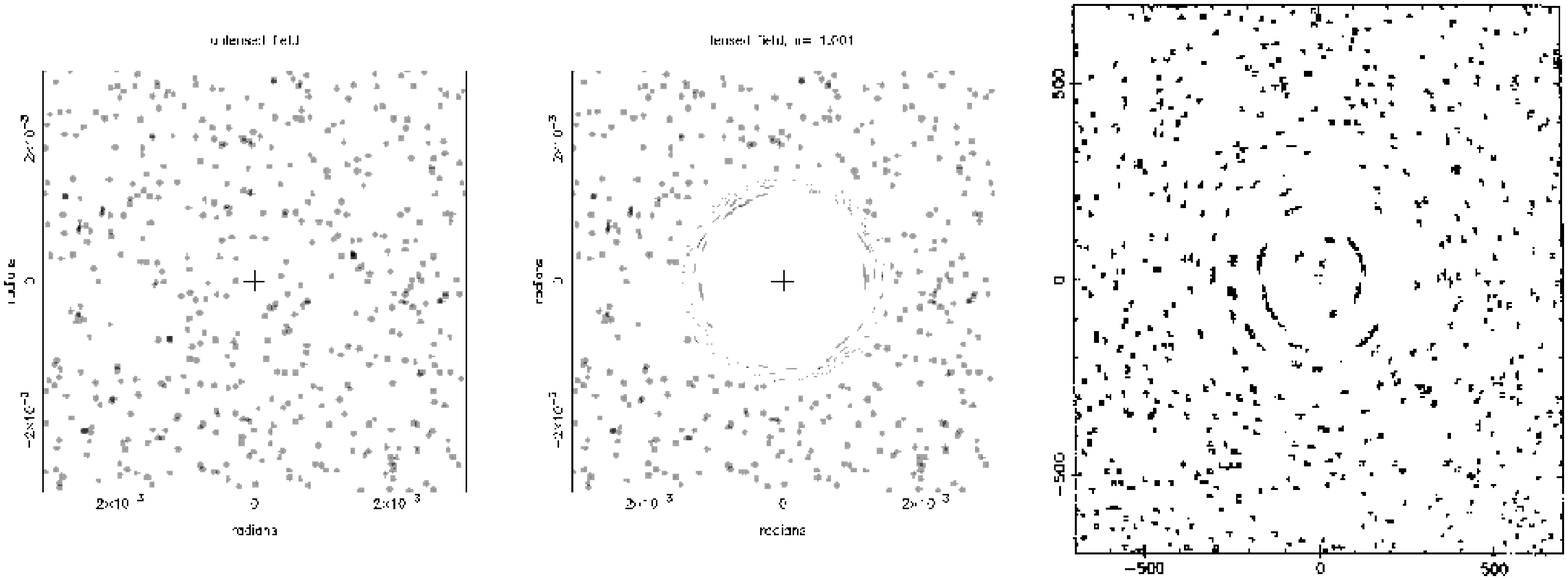,width=0.5\textwidth}}
\caption{Distortion field generated by the simulated gravity ball. The left
panel shows the grid of randomly distributed background sources as it
would be seen in the absence of the lens. The middle panel shows the
same population once they are distorted by a foreground gravity ball
with the parameters: $z_s=1$, $z_{l}=0.3$, $\theta_C=5'$ rad,
$n=1.001$. The right panel shows the same population distorted by a
foreground (invisible) circular cluster ($z_s=1.3$, $z_{l}=0.4$,
$\sigma =1000$ $kms^{-1}$). The simulation in the right panel is taken
from [15] and the units for both axis are in arcsecs.} 
\end{figure} 

\section{Conclusions.}
In the present paper we have discussed a special kind of nontopological
soliton, known as gravity ball, and examined the possibility for it to
be a gravitational lens. We investigated its lensing properties,
calculated the deflection angle of a point source, derived the lens
equation and plotted the lens curve for the gravity ball with specific
parameters.  We have shown that depending on the parameters of the
model there can be one, two or three images (two inside the Einstein
radius and one outside). However, the G-ball gravitational lens has
interesting features which are not shared by other known gravitational
lenses. In the case of a source, lens and observer located on the
optical axis together with the Einstein ring there is always an image
in the centre. Besides, the lensing geometry is radically
different---there is no effect outside the lens and thus, Einstein
radius is always less than radius of the ball. The refractive index of
the gravity ball contributes to the decrease in the radius of the
Einstein ring and increase of its thickness. For the extended sources
we showed how a large gravity ball can induce surprisingly large
distortions of images of distant galaxies. It is assumed that in order
to produce large arcs the cluster must have its surface-mass density
approximately supercritical, $\Sigma \ge \Sigma_{crit}$. We found that
the empty balls alone are able to produce arc-like features, including
straight arcs, arclets and radial arcs. We have modelled the
gravitational lensing effect of our gravity ball on a background source
field and found that the ball produces distortion (`shear') of that
field, consistent to some extent with the observations. We also
obtained constraints on the size of the large gravity ball, which can
be inferred from the existing observations of clusters with arcs. Empty
G-balls can show in observations as arcs without evident mass
distribution---a lens and/or through the distortion of the background
random galaxy field. Clearly, there are factors that this analysis has
not taken into account, for example, the gravity ball with the complex
mass distribution inside. These issues will be addressed in future papers.

Observationally empty G-balls have not yet been found, but we conclude from
our preliminary analysis that the existence of large G-balls cannot
be ruled out by gravitational lensing effects.

\section*{Acknowledgements}
MS is supported by a ICCR scholarship (Indo-Russian Exchange programme)
and acknowledges the hospitality of IUCAA, Pune. We would like to deeply 
thank Tarun Deep Saini for his expert assistance with the software.

\begin{table*}[p]
\caption{Lensing cluster sample}
\centerline{
\epsfig{figure=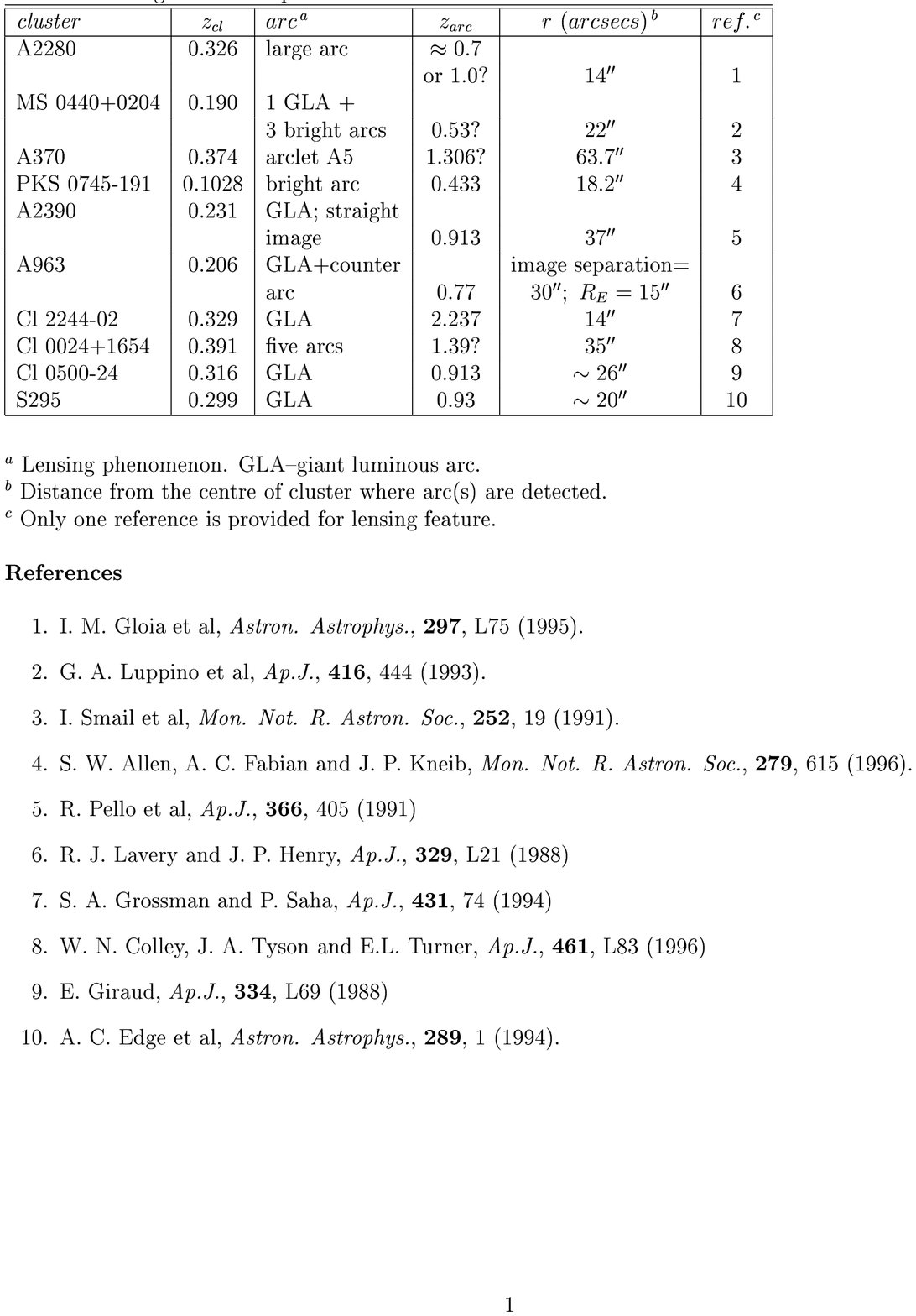}}
\end{table*}


\begin{thebibliography}{99}

\bibitem[1]{1} G. B. Gelmini, M. Gleiser and E. W. Kolb, {\it Phys. Rev. D.}, {\bf 39} (6), 1558 (1989). 
\bibitem[2]{2} B. Holdom, {\it Phys. Rev. D}, {\bf 36}, 1000 (1986).
\bibitem[3]{3} A. D. Dolgov and O. Yu. Markin, {\it Sov. Phys. JETP}, {\bf 71}, 207 (1990).
\bibitem[4]{4} S. Coleman, {\it Nucl. Phys.}, {\bf B262}, 263 (1985).
\bibitem[5]{5} T. D. Lee and Y. Pang, {\it Phys. Rev. D.}, {\bf 35}, 3678 (1987), and references therin.
\bibitem[6]{6} T. D. Lee and G. C. Wick, {\it Phys. Rev. D}, {\bf 9}, 229 (1974).
\bibitem[7]{7} R. Friedberg, T. D. Lee and A. Sirlin, {\it Phys. Rev. D.}, {\bf 13}, 2739 (1976); {\it Nucl. Phys.} {\bf B115}, 1 (1976); {\bf B115}, 32 (1976).
\bibitem[8]{8} M. Sethi and D. Lohiya, {\it Class. Quan. Grav.}, 1545, (1999).  
\bibitem[9]{9} A. Vilenkin, {\it Ap. J.}, {\bf L51}, 282 (1984).
\bibitem[10]{10} M. P. Dabrowski and F. E. Schunck, astro-ph/9807039.
\bibitem[11]{11} M. V. Sazhin et al, {\it Phys. Let. A.}, {\bf 215}, 199 (1996). \bibitem[12]{12} M. V. Sazhin, A. G. Yagola, A. V. Yakubov and A. F. Zakharov, {\it Astroph. Space Sci.}, {\bf 252}, 365 (1997). 
\bibitem[13]{13} P. Schneider, J. Ehlers and E. E. Falco, {\it in:} ``Gravitational lenses", 
Heidelberg: Springer 1992.
\bibitem[14]{14} A. F. Zakharov and M. V. Sazhin, {\it Physics--Uspekhi}, {\bf 41}, 945 (1998).
\bibitem[15]{15} B. Fort and Y. Mellier, {\it  Astron. Astroph. Rev.}, {\bf 5}, 239 (1994).
\end{thebibliography}
\end{document}